\documentclass{sig-alternate}
\usepackage{color,soul}
\usepackage{multirow}
\usepackage{url}
\usepackage{hhline}
\usepackage{dashrule}
\usepackage{booktabs}
\usepackage{array}
\usepackage{flushend}

\usepackage{breakurl}
\usepackage[breaklinks]{hyperref}

\begin{document}

\title{Making the Most of Tweet-Inherent Features for Social Spam Detection on Twitter}

\numberofauthors{1} 
\author{
\alignauthor
Bo Wang, Arkaitz Zubiaga, Maria Liakata, Rob Procter \\
       \affaddr{Department of Computer Science}\\
       \affaddr{University of Warwick}\\
       \affaddr{Coventry, UK}\\
       \email{\{bo.wang,a.zubiaga,m.liakata,rob.procter\}@warwick.ac.uk}
}

\date{25 January 2015}

\maketitle
\begin{abstract}
 Social spam produces a great amount of noise on social media services such as Twitter, which reduces the signal-to-noise ratio that both end users and data mining applications observe. Existing techniques on social spam detection have focused primarily on the identification of spam accounts by using extensive historical and network-based data. In this paper we focus on the detection of spam tweets, which optimises the amount of data that needs to be gathered by relying only on tweet-inherent features. This enables the application of the spam detection system to a large set of tweets in a timely fashion, potentially applicable in a real-time or near real-time setting. Using two large hand-labelled datasets of tweets containing spam, we study the suitability of five classification algorithms and four different feature sets to the social spam detection task. Our results show that, by using the limited set of features readily available in a tweet, we can achieve encouraging results which are competitive when compared against existing spammer detection systems that make use of additional, costly user features. Our study is the first that attempts at generalising conclusions on the optimal classifiers and sets of features for social spam detection over different datasets.
\end{abstract}


\vspace{1mm}
\noindent
{\bf General Terms:} Experimentation

\vspace{1mm}

\section{Introduction}
Social networking spam, or social spam, is increasingly affecting social networking websites, such as Facebook, Pinterest and Twitter. According to a study by the social media security firm Nexgate \cite{nguyen2013research}, social media platforms experienced a 355\% growth of social spam during the first half of 2013. Social spam can reach a surprisingly high visibility even with a simple bot \cite{DBLP:journals/corr/AielloDSR14}, which detracts from a company's social media presence and damages their social marketing ROI (Return On Investment). Moreover, social spam exacerbates the amount of unwanted information that average social media users receive in their timeline, and can occasionally even affect the physical condition of vulnerable users through the so-called ``\textit{Twitter psychosis}'' \cite{KalbitzerJ2014Twitter}.

Social spam has different effects and therefore its definition varies across major social networking websites. One of the most popular social networking services, Twitter, has published their definition of spamming as part of their ``The Twitter Rules'' \footnote{https://support.twitter.com/articles/18311-the-twitter-rules} and provided several methods for users to report spam such as tweeting ``\textit{@spam @username}'' where \textit{@username} will be reported as a spammer. While as a business, Twitter is also generous with mainline bot-level access \footnote{http://www.newyorker.com/tech/elements/the-rise-of-twitter-bots} and allows some level of advertisements as long as they do not violate ``The Twitter Rules''. In recent years we have seen Twitter being used as a prominent knowledge base for discovering hidden insights and predicting trends from finance to public sector, both in industry and academia. The ability to sort out the signal (or the information) from Twitter noise is crucial, and one of the biggest effects of Twitter spam is that it significantly reduces the signal-to-noise ratio. Our work on social spam is motivated by the initial attempts at harvesting a Twitter corpus around a specific topic with a set of predefined keywords \cite{zubiaga2015towards}. This led to the identification of a large amount of spam within those datasets. The fact that certain topics are trending and therefore many are tracking its contents encourages spammers to inject their spam tweets using the keywords associated with these topics to maximise the visibility of their tweets. These tweets produce a significant amount of noise both to end users who follow the topic as well as to tools that mine Twitter data.

In previous works, the automatic detection of Twitter spam has been addressed in two different ways. The first way is to tackle the task as a user classification problem, where a user can be deemed either a spammer or a non-spammer. This approach, which has been used by the majority of the works in the literature so far (see e.g., \cite{secryptWang10}, \cite{benevenuto@ceas10}, \cite{mccord2011detection}, \cite{Lee11sevenmonths}, \cite{Yang:2011:DFL:2186328.2186350} and \cite{DBLP:journals/corr/FerraraVDMF14}), makes use of numerous features that need to gather historical details about a user, such as tweets that a user posted in the past to explore what they usually tweet about, or how the number of followers and followings of a user has evolved in recent weeks to discover unusual behaviour. While this is ideal as the classifier can make use of extensive user data, it is often unfeasible due to restrictions of the Twitter API. The second, alternative way, which has not been as common in the literature (see e.g., \cite{benevenuto@ceas10}), is to define the task as a tweet classification problem, where a tweet can be deemed spam or non-spam. In this case, the classification task needs to assume that only the information provided within a tweet is available to determine if it has to be categorised as spam. Here, we delve into this approach to Twitter spam classification, studying the categorisation of a tweet as spam or not from its inherent features. While this is more realistic for our scenario, it presents the extra challenge that the available features are rather limited, which we study here.

In this work, after discussing the definition of social spam and reviewing previous research in Twitter spam detection, we present a comparative study of Twitter spam detection systems. We investigate the use of different features inherent to a tweet so as to identify the sets of features that do best in categorising tweets as spam or not. Our study compares five different classification algorithms over two different datasets. The fact that we test our classifiers on two different datasets, collected in different ways, enables us to validate the results and claim repeatability. Our results suggest a competitive performance can be obtained using tree-based classifiers for spam detection even with only tweet-inherent features, as comparing to the existing spammer detection studies. Also the combination of different features generally lead to an improved performance, with User feature + Bi \& Tri-gram (Tf) having the best results for both datasets.

\section{Social Spam}

The detection of spam has now been studied for more than a decade since email spam \cite{carreras2001boosting}. In the context of email messages, spam has been widely defined as ``unsolicited bulk email" \cite{blanzieri2008survey}. The term ``spam" has then been extended to other contexts, including ``social spam" in the context of social media. Similarly, social spam can be defined as the ``unwanted content that appears in online social networks". It is, after all, the noise produced by users who express a different behavior from what the system is intended for, and has the goal of grabbing attention by exploiting the social networks' characteristics, including for instance the injection of unrelated tweet content in timely topics, sharing malicious links or fraudulent information. Social spam hence can appear in many different forms, which poses another challenge of having to identify very different types of noise for social spam detection systems.

\subsection{Social Spammer Detection}

As we said before, most of the previous work in the area has focused on the detection of users that produce spam content (i.e., spammers), using historical or network features of the user rather than information inherent to the tweet. Early work by \cite{secryptWang10}, \cite{benevenuto@ceas10} and \cite{mccord2011detection} put together a set of different features that can be obtained by looking at a user's previous behaviour. These include some aggregated statistics from a user's past tweets such as average number of hashtags, average number of URL links and average number of user mentions that appear in their tweets. They combine these with other non-historical features, such as number of followers, number of followings and age of the account, which can be obtained from a user's basic metadata, also inherent to each tweet they post. Some of these features, such as the number of followers, can be gamed by purchasing additional followers to make the user look like a regular user account.

Lee et al. \cite{Lee11sevenmonths} and Yang et al. \cite{Yang:2011:DFL:2186328.2186350} employed different techniques for collecting data that includes spam (more details will be discussed in Section \ref{ssec:datasets}) and performed comprehensive studies of the spammers' behaviour. They both relied on the tweets posted in the past by the users and their social networks, such as tweeting rate, following rate, percentage of bidirectional friends and local clustering coefficient of its network graph, aiming to combat spammers' evasion tactics as these features are difficult or costly to simulate. Ferrara et al. \cite{DBLP:journals/corr/FerraraVDMF14} used network, user, friends, timing, content and sentiment features for detecting Twitter bots, their performance evaluation is based on the social honeypots dataset (from \cite{Lee11sevenmonths}). Miller et al. \cite{Miller201464} treats spammer detection as an anomaly detection problem as clustering algorithms are proposed and such clustering model is built on normal Twitter users with outliers being treated as spammers. They also propose using 95 uni-gram counts along with user profile attributes as features. The sets of features utilised in the above works require the collection of historical and network data for each user, which do not meet the requirements of our scenario for spam detection.

\subsection{Social Spam Detection}

Few studies have addressed the problem of spam detection. Santos et al. \cite{santosCISIS2013} investigated two different approaches, namely compression-based text classification algorithms (i.e. Dynamic Markov compression and Prediction by partial matching) and using ``bag of words'' language model (also known as uni-gram language model) for detecting spam tweets. Martinez-Romo and Araujo \cite{MartinezRomo20132992} applied Kullback-Leibler Divergence and examined the difference of language used in a set of tweets related to a trending topic, suspicious tweets (i.e. tweets that link to a web page) and the page linked by the suspicious tweets. These language divergence measures were used as their features for the classification. They used several URL blacklists for identifying spam tweets from their crawled dataset, therefore each one of their labelled spam tweets contains a URL link, and is not able to identify other types of spam tweets. In our studies we have investigated and evaluated the discriminative power of four feature sets on two Twitter datasets (which were previously in \cite{Lee11sevenmonths} and \cite{Yang:2011:DFL:2186328.2186350}) using five different classifiers. We examine the suitability of each of the features for the spam classification purposes. Comparing to \cite{MartinezRomo20132992} our system is able to detect most known types of spam tweet irrespective of having a link or not. Also our system does not have to analyze a set of tweets relating to each topic (which \cite{MartinezRomo20132992} did to create part of their proposed features) or external web page linked by each suspicious tweet, therefore its computation cost does not increase dramatically when applied for mass spam detection with potentially many different topics in the data stream. 

The few works that have dealt with spam detection are mostly limited in terms of the sets of features that they studied, and the experiments have been only conducted in a single dataset (except in the case of \cite{MartinezRomo20132992}, where very limited evaluation was conducted on a new and smaller set of tweets), which does not allow for generalisability of the results. To the best of our knowledge, our work is the first study that evaluates a wide range of tweet-inherent features (namely user, content, n-gram and sentiment features) over two different datasets, obtained from \cite{Lee11sevenmonths} and \cite{Yang:2011:DFL:2186328.2186350} and with more than 10,000 tweets each, for the task of spam detection. The two datasets were collected using completely different approaches (namely deploying social honeypots for attracting spammers; and checking malicious URL links), which helps us learn more about the nature of social spam and further validate the results of different spam detection systems.

\section{Methodology}
In this section we describe the Twitter spam datasets we used, the text preprocessing techniques that we performed on the tweets, and the four different feature sets we used for training our spam vs non-spam classifier.

\subsection{Datasets}
\label{ssec:datasets}
A labelled collection of tweets is crucial in a machine learning task such as spam detection. We found no spam dataset which is publicly available and specifically fulfils the requirements of our task. Instead, the datasets we obtained include Twitter users labelled as spammers or not. For our work, we used the latter, which we adapted to our purposes by taking out the features that would not be available in our scenario of spam detection from tweet-inherent features. We used two spammer datasets in this work, which have been created using different data collection techniques and therefore is suitable to our purposes of testing the spam classifier in different settings. To accomodate the datasets to our needs, we sample one tweet for each user in the dataset, so that we can only access one tweet per user and cannot aggregate several tweets from the same user or use social network features. In what follows we describe the two datasets we use.

\textbf{Social Honeypot Dataset:} Lee et al. \cite{Lee11sevenmonths} created and manipulated (by posting random messages and engaging in none of the activities of legitimate users) 60 social honeypot accounts on Twitter to attract spammers. Their dataset consists of 22,223 spammers and 19,276 legitimate users along with their most recent tweets. They used Expectation-Maximization (EM) clustering algorithm and then manually grouped their harvested users into 4 categories: duplicate spammers, duplicate @ spammers, malicious promoters and friend infiltrators. \textbf{1KS-10KN Dataset:} Yang et al. \cite{Yang:2011:DFL:2186328.2186350} defines a tweet that contains at least one malicious or phishing URL as a spam tweet, and a user whose spam ratio is higher than 10\% as a spammer. Therefore their dataset which contains 1,000 spammers and 10,000 legitimate users, represents only one major type of spammers (as discussed in their paper).

We used \textit{spammer vs. legitimate user} datasets from \cite{Lee11sevenmonths} and \cite{Yang:2011:DFL:2186328.2186350}. After removing duplicated users and the ones that do not have any tweets in the dataset we randomly selected one tweet from each spammer or legitimate user to create our labelled collection of \textit{spam vs. legitimate tweets}, in order to avoid overfitting and reduce our sampling bias. The resulting datasets contain 20,707 spam tweets and 19,249 normal tweets (named Social Honeypot dataset, as from \cite{Lee11sevenmonths}), and 1,000 spam tweets and 9,828 normal tweets (named 1KS-10KN dataset, as from \cite{Yang:2011:DFL:2186328.2186350}) respectively. 

\subsection{Data Preprocessing}
Before we extract the features to be used by the classifier from each tweet, we apply a set of preprocessing techniques to the content of the tweets to normalise it and reduce the noise in the classification phase. The preprocessing techniques include decoding HTML entities, and expanding contractions with apostrophes to standard spellings (e.g. ``I'm'' -> ``I am''). More advanced preprocessing techniques such as spell-checking and stemming were tested but later discarded given the minimal effect we observed in the performance of the classifiers.

For the specific case of the extraction of sentiment-based features, we also remove hashtags, links, and user mentions from tweet contents.

\subsection{Features}
As spammers and legitimate users have different goals in posting tweets or interacting with other users on Twitter, we can expect that the characteristics of spam tweets are quite different to the normal tweets. The features inherent to a tweet include, besides the tweet content itself, a set of metadata including information about the user who posted the tweet, which is also readily available in the stream of tweets we have access to in our scenario. We analyse a wide range of features that reflect user behaviour, which can be computed straightforwardly and do not require high computational cost, and also describe the linguistic properties that are shown in the tweet content. We considered four feature sets: (i) user features, (ii) content features, (iii) n-grams, and (iv) sentiment features.

\begin{table*}[htb]
 \begin{center}
  \begin{tabular}{ l | l }
    \toprule[1pt]
    \multicolumn{1}{c |}{\textbf{User features}} & \multicolumn{1}{c}{\textbf{Content features}} \\
    \midrule[.5pt]
    Length of profile name  &  Number of words  \\
    \midrule[.5pt]
    Length of profile description  &  Number of characters  \\
    \midrule[.5pt]
    Number of followings (FI)  &  Number of white spaces  \\
    \midrule[.5pt]
    Number of followers (FE)  &  Number of capitalization words  \\
    \midrule[.5pt]
    Number of tweets posted  &  Number of capitalization words per word  \\
    \midrule[.5pt]
    Age of the user account, in hours (AU)  &  Maximum word length  \\
    \midrule[.5pt]
    Ratio of number of followings and followers (FE/FI)  &  Mean word length  \\
    \midrule[.5pt]
    Reputation of the user (FE/(FI + FE))  &  Number of exclamation marks \\
    \midrule[.5pt]
    Following rate (FI/AU)  &  Number of question marks  \\
    \midrule[.5pt]
    Number of tweets posted per day  &  Number of URL links  \\
    \midrule[.5pt]
    Number of tweets posted per week  &  Number of URL links per word  \\
    \midrule[.5pt]
    \multicolumn{1}{c |}{\textbf{N-grams}} &  Number of hashtags  \\
    \midrule[.5pt]
    Uni + bi-gram or bi + tri-gram  &  Number of hashtags per word  \\
    \midrule[.5pt]
      &  Number of mentions  \\
    \midrule[.5pt]
    \multicolumn{1}{c |}{\textbf{Sentiment features}}  &  Number of mentions per word  \\
    \midrule[.5pt]
     Automatically created sentiment lexicons  &  Number of spam words  \\
    \midrule[.5pt]
     Manually created sentiment lexicons &  Number of spam words per word \\
    \midrule[.5pt]
      &  Part of speech tags of every tweet \\
    \bottomrule[1pt]
  \end{tabular}
 \end{center}
 \caption{List of features}
 \label{tab:featurelist}
\end{table*}

\textbf{User features} include a list of 11 attributes about the author of the tweet (as seen in Table \ref{tab:featurelist}) that is generated from each tweet's metadata, such as reputation of the user \cite{secryptWang10}, which is defined as the ratio between the number of followers and the total number of followers and followings and it had been used to measure user influence. Other candidate features, such as the number of retweets and favourites garnered by a tweet, were not used given that it is not readily available at the time of posting the tweet, where a tweet has no retweets or favourites yet.

\textbf{Content features} capture the linguistic properties from the text of each tweet (Table \ref{tab:featurelist}) including a list of content attributes and part-of-speech tags. Among the 17 content attributes, number of spam words and number of spam words per word are generated by matching a popular list of spam words \footnote{https://github.com/splorp/wordpress-comment-blacklist/blob/master/blacklist.txt}. Part-of-speech (or POS) tagging provides syntactic (or grammatical) information of a sentence and has been used in the natural language processing community for measuring text informativeness (e.g. Tan et al. \cite{DBLP:journals/corr/TanLP14} used POS counts as a informativeness measure for tweets). We have used a Twitter-specific tagger \cite{Gimpel:2011:PTT:2002736.2002747}, and in the end our POS feature consists of uni-gram and 2-skip-bi-gram representations of POS tagging for each tweet in order to capture the structure and therefore informativeness of the text. We also used Stanford tagger with standard Penn Tree tags, which makes very little difference in the classification results.

\textbf{N-gram models} have long been used in natural language processing for various tasks including text classification. Although it is often criticized for its lack of any explicit representation of long range or semantic dependency, it is surprisingly powerful for simple text classification with reasonable amount of training data. In order to give the best classification result while being computationally efficient we have tried uni + bi-gram or bi + tri-gram with binary (i.e. 1 for feature presence while 0 for absence), term-frequency (tf) and tf-idf (i.e. Term Frequency times Inverse Document Frequency) techniques.

\textbf{Sentiment features:} Ferrara et al. \cite{DBLP:journals/corr/FerraraVDMF14} used tweet-level sentiment as part of their feature set for the purpose of detecting Twitter bots. We have used the same list of lexicons from \cite{MohammadKZ2013} (which has been proved of achieving top performance in the Semeval-2014 Task 9 Twitter sentiment analysis competition) for generating our sentiment features, including manually generated sentiment lexicons: AFINN lexicon \cite{nielsen2011new}, Bing Liu lexicon \cite{liu2010sentiment}, MPQA lexicon \cite{Wilson:2005:RCP:1220575.1220619}; and automatically generated sentiment lexicons: NRC Hashtag Sentiment lexicon \cite{MohammadKZ2013} and Sentiment140 lexicon \cite{MohammadKZ2013}.

\section{Evaluation}

\subsection{Selection of Classifier}
During the classification and evaluation stage, we tested 5 classification algorithms implemented using scikit-learn\footnote{http://scikit-learn.org/}: Bernoulli Naive Bayes, K-Nearest Neighbour (KNN), Support Vector Machines (SVM), Decision Tree, and Random Forests. These algorithms were chosen as being the most commonly used in the previous research on spammer detection. We evaluate using the standard information retrieval metrics of recall (R), precision (P) and F1-measure. Recall in this case refers to the ratio obtained from diving the number of correctly classified spam tweets (i.e. True Positives) by the number of tweets that are actually spam (i.e. True Positives + False Negatives). Precision is the ratio of the number of correctly classified spam tweets (i.e. True Positives) to the total number of tweets that are classified as spam (i.e. True Positives + False Positives). F1-measure can be interpreted as a harmonic mean of the precision and recall, where its score reaches its best value at 1 and worst at 0. It is defined as: 
\begin{displaymath}
F1 = 2 * (precision * recall) / (precision + recall)
\end{displaymath}

In order to select the best classifier for our task, we have used a subset of each dataset (20\% for 1KS-10KN dataset and 40\% for Social Honeypot dataset, due to the different sizes of the two datasets) to run a 10-fold cross validation for optimising the hyperparameters of each classifier. By doing so it minimises the risk of over-fitting in model selection and hence subsequent selection bias in performance evaluation. Such optimisation was conducted using all 4 feature sets (each feature was normalised to fit the range of values [-1, 1]; we also selected 30\% of the highest scoring features using Chi Square for tuning SVM as computationally it is more efficient and gives better classification results). Then we evaluated our algorithm on the rest of the data (i.e. 80\% for 1KS-10KN dataset and 60\% for Social Honeypot dataset), again using all 4 feature sets in a 10-fold cross validation setting (same as in grid-search, each feature was normalised and Chi square feature selection was used for SVM).

As shown in Table \ref{tab:classifiers}, tree-based classifiers achieved very promising performances, among which Random Forests outperform all the others when we look at the F1-measure. This outperformance occurs especially due to the high precision values of 99.3\% and 94.1\% obtained by the Random Forest classifier. While Random Forests show a clear superiority in terms of precision, its performance in terms of recall varies for the two datasets; it achieves high recall for the Social Honeypot dataset, while it drops substantially for the 1KS-10KN dataset due to its approximate 1:10 spam/non-spam ratio. These results are consistent with the conclusion of most spammer detection studies; our results extend this conclusion to the spam detection task.

When we compare the performance values for the different datasets, it is worth noting that with the Social Honeypot dataset the best result is more than 10\% higher than the best result in 1KS-10KN dataset. This is caused by the different spam/non-spam ratios in the two datasets, as the Social Honeypot dataset has a roughly 50:50 ratio while in 1KS-10KN it is roughly 1:10 which is a more realistic ratio to reflect the amount of spam tweets existing on Twitter (In Twitter's 2014 Q2 earnings report it says that less than 5\% of its accounts are spam\footnote{http://www.webcitation.org/6VyBTJ7vt}, but independent researchers believe the number is higher). In comparison to the original papers, \cite{Lee11sevenmonths} reported a best 0.983 F1-score and \cite{Yang:2011:DFL:2186328.2186350} reported a best 0.884 F1-score. Our results are only about 4\% lower than their results, which make use of historical and network-based data, not readily available in our scenario. Our results suggest that a competitive performance can also be obtained for spam detection where only tweet-inherent features can be used.

\begin{table*}[bth]
\centering
\begin{tabular}{ |c|c|c|c|c|c|c|  }
 \hline
  \multicolumn{1}{|c|}{\multirow{2}{*}{\textbf{Classifier}} } &
  \multicolumn{3}{|c|}{\textbf{1KS-10KN Dataset}} &
  \multicolumn{3}{|c|}{\textbf{Social Honeypot Dataset}} \\
  
  \multicolumn{1}{|c|}{} &
  \multicolumn{1}{|c|}{\textbf{Precision}} &
  \multicolumn{1}{|c|}{\textbf{Recall}} &
  \multicolumn{1}{|c|}{\textbf{F-measure}} &
  \multicolumn{1}{|c|}{\textbf{Precision}} &
  \multicolumn{1}{|c|}{\textbf{Recall}} &
  \multicolumn{1}{|c|}{\textbf{F-measure}} \\
 \hline
 Bernoulli NB   & 0.899 & 0.688 & 0.778      & 0.772 & 0.806 & 0.789 \\
 KNN            & 0.924 & 0.706 & 0.798      & 0.802 & 0.778 & 0.790 \\
 SVM            & 0.872 & 0.708 & 0.780      & 0.844 & 0.817 & 0.830 \\
 Decision Tree  & 0.788 & 0.782 & 0.784      & 0.914 & 0.916 & 0.915 \\
 Random Forest  & 0.993 & 0.716 & \bf{0.831} & 0.941 & 0.950 & \bf{0.946} \\
 \hline
\end{tabular}
\caption{Comparison of performance of classifiers}
\label{tab:classifiers}
\end{table*}

\subsection{Evaluation of Features}
We trained our best classifier (i.e. Random Forests) with different feature sets, as well as combinations of the feature sets using the two datasets (i.e. the whole corpora), and under a 10-fold cross validation setting. We report our results in Table \ref{tab:features}. As seen in 1KS-10KN dataset, the F1-measure for different feature sets ranges from 0.718 to 0.820 when using a single feature set. All feature set combinations except C + S (content + sentiment feature) perform higher than 0.810 in terms of F1-measure, reflecting that feature combinations have more discriminative power than a single feature set. 

For the Social Honeypot dataset, we can clearly see User features (U) having the most discriminative power as it has a 0.940 F1-measure. Results without using User features (U) have significantly worse performance, and feature combinations with U give very little improvement with respect to the original 0.940 (except for U + Uni \& Bi-gram (Tf) + S). This means U is dominating the discriminative power of these feature combinations and other feature sets contribute very little in comparison to U. This is potentially caused by the data collection approach (i.e. by using social honeypots) adopted by \cite{Lee11sevenmonths}, which resulted in the fact that most spammers that they attracted have distinguishing user profile information compared to the legitimate users. On the other hand, Yang et al. \cite{Yang:2011:DFL:2186328.2186350} checked malicious or phishing URL links for collecting their spammer data, and this way of data collection gives more discriminative power to Content and N-gram features than \cite{Lee11sevenmonths} does (although U is still a very significant feature set in 1KS-10KN). Note that U + Bi \& Tri-gram (Tf) resulted in the best performance in both datasets, showing that these two feature sets are the most beneficial to each other irrespective of the different nature of datasets.

Another important aspect to take into account when choosing the features to be used is the computation time, especially when one wants to apply the spam classifier in real-time. Table \ref{tab:efficiency} shows a efficiency comparison for generating each feature from 1000 tweets, using a machine with 2.8 GHz Intel Core i7 processor and 16 GB memory. Some of the features, such as the User features, can be computed quickly and require minimal computational cost, as most of these features can be straightforwardly inferred from a tweet's metadata. Other features, such as N-grams and part-of-speech counts (from Content features), can be affected by the size of the vocabulary in the training set. On the other hand, some of the features are computationally more expensive, and therefore worth studying their applicability. This is the case of Sentiment features, which require string matching between our training documents and a list of lexica we used. We keep the sentiment features since they have shown added value in the performance evaluation of feature set combinations. Similarly, Content features such as \textit{Number of spam words} and \textit{Number of spam words per word} also require string matching between our training documents and a dictionary containing 11,529 spam words. However, given that the latter did not provide significant improvements in terms of accuracy, most probably because the spam words were extracted from blogs, we conclude that \textit{Number of spam words} and \textit{Number of spam words per word} can be taken out from the representation for the sake of the classifier's efficiency.

\begin{table*}[bth]
\centering
\begin{tabular}{ |c|c|c|c|c|c|c|  }
 \hline
  \multicolumn{1}{|c|}{\multirow{2}{*}{\textbf{Feature Set}} } &
  \multicolumn{3}{|c|}{\textbf{1KS-10KN Dataset}} &
  \multicolumn{3}{|c|}{\textbf{Social Honeypot Dataset}} \\
  
  \multicolumn{1}{|c|}{} &
  \multicolumn{1}{|c|}{\textbf{Precision}} &
  \multicolumn{1}{|c|}{\textbf{Recall}} &
  \multicolumn{1}{|c|}{\textbf{F-measure}} &
  \multicolumn{1}{|c|}{\textbf{Precision}} &
  \multicolumn{1}{|c|}{\textbf{Recall}} &
  \multicolumn{1}{|c|}{\textbf{F-measure}} \\
 \hline
 User features (U)   & 0.895    & 0.709 & 0.791   & 0.938 & 0.940 & 0.940\\
 Content features (C) & 0.951  & 0.657 & 0.776   & 0.771 & 0.753 & 0.762\\
 Uni + Bi-gram (Binary) & 0.930 & 0.725 & 0.815   & 0.759 & 0.727 & 0.743\\
 Uni + Bi-gram (Tf) & 0.959 & 0.715 & 0.819   & 0.783 & 0.767 & 0.775\\
 Uni + Bi-gram (Tfidf) & 0.943 & 0.726 & 0.820   & 0.784 & 0.765 & 0.775\\
 Bi + Tri-gram (Tfidf) & 0.931 & 0.684 & 0.788   & 0.797 & 0.656 & 0.720\\
 Sentiment features (S)    & 0.966 & 0.574 & 0.718    & 0.679 & 0.727 & 0.702\\
  \hline
 U + C & 0.974  & 0.708 & 0.819   & 0.938 & 0.949 & \bf{0.943}\\
 U + Bi \& Tri-gram (Tf) & 0.972 & 0.745 & \bf{0.843}   & 0.937 & 0.949 & \bf{0.943}\\
 U + S & 0.948 & 0.732 & 0.825   & 0.940 & 0.944 & 0.942\\
 Uni \& Bi-gram (Tf) + S & 0.964 & 0.721 & 0.824   & 0.797 & 0.744 & 0.770\\
 C + S & 0.970 & 0.649 & 0.777   & 0.778 & 0.762 & 0.770\\
 C + Uni \& Bi-gram (Tf) & 0.968 & 0.717 & 0.823   & 0.783 & 0.757 & 0.770\\
  \hline
 U + C + Uni \& Bi-gram (Tf) & 0.985 & 0.727 & 0.835   & 0.934 & 0.949 & 0.941\\
 U + C + S  & 0.982 & 0.704 & 0.819   & 0.937 & 0.948 & 0.942\\
 U + Uni \& Bi-gram (Tf) + S  & 0.994 & 0.720 & 0.834   & 0.928 & 0.946 & 0.937\\
 C + Uni \& Bi-gram (Tf) + S  & 0.966 & 0.720 & 0.824   & 0.806 & 0.758 & 0.782\\
 \hline
 U + C + Uni \& Bi-gram (Tf) + S  & 0.988 & 0.725 & 0.835   & 0.936 & 0.947 & 0.942\\
 \hline
\end{tabular}
\caption{Performance evaluation of various feature set combinations}
\label{tab:features}
\end{table*} 

\begin{table*}[!bth]
\centering
\begin{tabular}{ |c|c|c|c|  }
 \hline
  \multicolumn{1}{|c|}{\multirow{2}{*}{\textbf{Feature set}} } &
  \multicolumn{1}{|c|}{\multirow{1}{*}{\textbf{Computation time (in seconds)}} } \\
  
  \multicolumn{1}{|c|}{} &
  \multicolumn{1}{|c|}{\textbf{for 1000 tweets}} \\
 \hline
 User features   & 0.0057 \\
 N-gram            & 0.3965 \\
 Sentiment features            & 20.9838 \\
 Number of spam words (NSW)  & 19.0111 \\
 Part-of-speech counts (POS) & 0.6139 \\
 Content features including NSW and POS  & 20.2367 \\
 Content features without NSW  & 1.0448 \\
 Content features without POS  & 19.6165 \\
 \hline
\end{tabular}
\caption{Feature engineering computation time for 1000 tweets}
\label{tab:efficiency}
\end{table*}

\section{Discussion}

Our study looks at different classifiers and feature sets over two spam datasets to pick the settings that perform best. First, our study on spam classification buttresses previous findings for the task of spammer classification, where Random Forests were found to be the most accurate classifier. Second, our comparison of four feature sets reveals the features that, being readily available in each tweet, perform best in identifying spam tweets. While different features perform better for each of the datasets when using them alone, our comparison shows that the combination of different features leads to an improved performance in both datasets. We believe that the use of multiple feature sets increases the possibility to capture different spam types, and makes it more difficult for spammers to evade all feature sets used by the spam detection system. For example spammers might buy more followers to look more legitimate but it is still very likely that their spam tweet will be detected as its tweet content will give away its spam nature.

Due to practical limitations, we have generated our spam vs. non-spam data from two spammer vs. non-spammer datasets that were collected in 2011. For future work, we plan to generate a labelled spam/non-spam dataset which was crawled in 2014. This will not only give us a purpose-built corpus of spam tweets to reduce the possible effect of sampling bias of the two datasets that we used, but will also give us insights on how the nature of Twitter spam changes over time and how spammers have evolved since 2011 (as spammers do evolve and their spam content are manipulated to look more and more like normal tweet). Furthermore we will investigate the feasibility of cross-dataset spam classification using domain adaptation methods, and also whether unsupervised approaches work well enough in the domain of Twitter spam detection.

A caveat of the approach we relied on for the dataset generation is the fact that we have considered spam tweets posted by users who were deemed spammers. This was done based on the assumption that the majority of social spam tweets on Twitter are shared by spam accounts. However, the dataset could also be complemented with spam tweets which are occasionally posted by legitimate users, which our work did not deal with. An interesting study to complement our work would be to look at these spam tweets posted by legitimate users, both to quantify this type of tweets, as well as to analyse whether they present different features from those in our datasets, especially when it comes to the user-based features as users might have different characteristics. For future work, we plan to conduct further evaluation on how our features would function for spam tweets shared by legitimate users, in order to fully understand the effects of bias of pursuing our approach of corpus construction.

\section{Conclusion}

In this paper we focus on the detection of spam tweets, solely making use of the features inherent to each tweet. This differs from most previous research works that classified Twitter users as spammers instead, and represents a real scenario where either a user is tracking an event on Twitter, or a tool is collecting tweets associated with an event. In these situations, the spam removal process cannot afford to retrieve historical and network-based features for all the tweets involved with the event, due to high number of requests to the Twitter API that this represents. We have tested five different classifiers, and four different feature sets on two Twitter spam datasets with different characteristics, which allows us to validate our results and claim repeatability. While the task is more difficult and has access to fewer data than a spammer classification task, our results show competitive performances. Moreover, our system can be applied for detecting spam tweets in real time and does not require any feature not readily available in a tweet.

Here we have conducted the experiments on two different datasets which were originally collected in 2011. While this allows us to validate the results with two datasets collected in very different methods, our plan for future work includes the application of the spam detection system to more recent events, to assess the validity of the classifier with recent data as Twitter and spammers may have evolved.

\bibliographystyle{abbrv}
\bibliography{spamdetection201501_ref} 

\balancecolumns
\end{document}